\documentstyle[12pt]{article}
\newcommand{\bfvec}[1] {\mbox{} {\mbox{}} {\mbox{\boldmath$ #1$}}}
\title{\small VIOLATION OF THE ENERGY CONSERVATION LAW IN LORENTZ-DIRAC
EQUATIONS FOR MORE THAN ONE CHARGE}

\author{\small D. Villarroel and R. Rivera  \\
{\small Departamento de F\'\i sica, } \\ {\rm \small Universidad T\'ecnica
Federico Santa Mar\'\i a}\\ {\rm\small Casilla 110-V, Valpara\'\i so, Chile}}
\date{}

\begin{document}
\maketitle

\normalsize

\setlength{\baselineskip}{1.5 \baselineskip}{

An exact solution of Lorentz-Dirac equations where the energy conservation law
is violated, is described herein for the case of two charges.

\vspace{4.5cm}


\newpage
Lorentz-Dirac equations are currently accepted as the equations of motion for
charged particles in classical electrodynamics.  These equations were derived
by Dirac in a classical paper in 1938 \cite{1}; the same equations were also
obtained by Rohrlich on the basis of an action principle in his well-known
book \cite{2}. In the case of two charges, the equations are:
\begin{equation}
m_1 a_1^\mu = \left(e_1/c\right)F_{{\rm ext}}^{\mu\alpha} v_\alpha^1 +
\left( e_1 / c \right) F_{{\rm 2ret}}^{\mu\alpha} v_\alpha^1 +
\left(2e_1^2/3c^3\right)\left( \dot{a}_1^{\mu} - \left( 1/c^2 \right)
a_1^\lambda a_\lambda^1 v_1^\mu \right)
\end{equation}
where we use Rohrlich's notation. The equation for charge $e_2$ is the same as
(1), but with indexes 1 and 2 interchanged. The first term on the right-hand
side (RHS) in (1), is Lorentz's force due to the external field. The second
term on the RHS, links charge $e_2$ charge $e_2$ by means of a purely retarded
interaction, and represents the only mutual interaction of the charged
particles in Lorentz-Dirac equations. Finally, the third term is the well
known radiation reaction term, constituted by two parts: ``Schott term",
involving the third derivative of the charge position, and ``Larmor term",
involving the square of the acceleration. In the case of only one charge in an
external field, the second term on the RHS of (1) disappears.

Equations in (1) are the final achievement of Dirac's conception, according to
which it is possible to formulate a consistent set of classical equations of
motion for point charges. Dirac reached these equations starting from Maxwell
equations and the principles of relativity, masking the divergences associated
with the singularities of the field at the position of the charges, by means
of a mass renormalization procedure. Extensive theoretical work, carried out
especially by Rohrlich \cite{2}, led to consider that Lorentz-Dirac equations
coherently describe the one-charge case in an external field.  However, the
situation does not remain the same when several charges are involved. In fact,
a few years after Dirac derived his equations, Eliezer \cite{3} applied them
to the case of two equal-mass particles with charges of equal magnitude but
opposite signs, moving in a straight line and colliding head-on. This led to a
controversy that still holds true \cite{4}.  According to Eliezer, the charges
stop before colliding, then turn back and move away from each other, with an
acceleration that is always different from zero, and with a velocity that
tends toward the speed of light. This behavior is contrary to what common
understanding would suggest for two charges of opposite signs, and is also
contradictory to the law of energy conservation. Eliezer's conclusions were
questioned by Clavier
\cite{5}, Plass \cite{6}, and Rohrlich \cite{2}. However, more
recent detailed numerical calculations by Baylis and Huschilt
\cite{7}, and Kasher
\cite{8}, came to support Eliezer's conclusions.

The pathological behavior of Lorentz-Dirac equations in Eliezer head-on
collision has received, in general, little attention in textbooks (one
exception is Parrott's book \cite{4} ). Several reasons may help to explain
this lack of attention. In spite of the one-dimensional character of Eliezer's
collision, the retarded nature of the interaction between the charges gives
rise to mathematical complications. In particular, the numerical treatment of
\cite{7}  \& \cite{8} are rather elaborated, and
questions about the convergence of the iterative process cannot be easily
answered. This problem is related, in turn, to the issue of the existence and
the uniqueness of the solution of Lorentz-Dirac equations, which still remains
an open question in the case of several charges. Another source of troubles
comes from the fact that in Eliezer's collision, the kinetic energy of the
charges, the total rate of the radiation emitted by them, as well as the
energy stored in the field of the charges, are all time-dependent. This makes
it very difficult to analyze the energy conservation law at any given moment.
Although the kinetic energy and the total rate of the radiation are well
defined concepts, the same does not occur with the concept of ``energy stored
in the field of the charges", which makes for even further troubles. We also
want to point out that the catastrophic behavior of Eliezer's solutions has
been obtained merely as a mathematical result of Lorentz-Dirac equations, but
without identifying any physical cause- an identification that would be
convenient for a deeper understanding.

 From our point of view, Lorentz-Dirac equations cannot be correct, because
they do not take into account all the radiation emitted by the charges. In
fact, in the case of two charges, radiation must be described by: a Larmor
term for each charge, terms that are indeed present in (1); plus a term that
considers the interference effect between the fields of both charges, which is
not present in (1). The second term on the RHS of (1) is the only one that
considers both particles, but this term has very little to do with the
interference radiation, which must involve the product of the acceleration of
two charges, and this is clearly not the case. So, in order to put this
conjecture in a quantitative way, it would be ideal to find one exact solution
of the Lorentz-Dirac equations for two charges where the total rate of
radiation (including the interference term) can be calculated in a precise
way. It is the purpose of this note to present one such solution, which is the
first exact solution for the Lorentz-Dirac equations for more than one charge
present in the literature related to the subject \cite{9} . As it will be
shown below, Lorentz-Dirac equations, with appropriate external fields to be
determined, allow for the motion of two charges moving in a plane at opposite
ends of a diameter revolving at constant angular velocity $\omega$ in a fixed
circular orbit of radius $a$.  This solution can be considered as a sort of
generalization of the exact solution found by Sokolov and Kolesnikova
\cite{10}  for one charge
in a circular orbit.

Before discussing the algebraic treatment of our solution, it may be
convenient to make some general comments on Lorentz-Dirac equations, since
treatment of this subject is almost absent in current literature. Besides,
they may help those readers not very familiar with radiation reaction effects,
and, on the other hand, they will allow us to set the framework of our
discussion.

It may seem rather natural to study the validity of Lorentz-Dirac equations by
means of experiments, instead of resorting to theoretical disquisition as we
are doing here. However, due to other effects that appear in experiments that
make it almost impossible to reach a clear conclusion, the troubles with
Lorentz-Dirac equations cannot be clarified with the help of practical
measurements. The motion of an electron in a storage ring is a good example
for illustrating this situation, since in these machines radiation reaction
plays an important role. The starting point for the description of the
electron motion in a storage ring is Lorentz-Dirac equations for one electron
\cite{11,12}. Now, in a
machine of a few Gev, the magnetic external force is much greater than the
Larmor term, which in turn is much greater than the Schott term. These
differences make it difficult to test Lorentz-Dirac equations; in addition,
there are several other complications. For example, in the vacuum chamber of
the machine there is not only one electron, but many of them interacting with
each other. In particular, these interactions lead to a loss of electrons from
the beam. Furthermore, the particle trajectory is influenced by the
interaction of the electron with the wall of the vacuum chamber, as well as by
the inhomogeneity of the magnetic field. But the most striking phenomenon that
rules out the use of Lorentz-Dirac equations for an accurate description of
the motion are well-established quantum effects on electron motion
\cite{11,12}.

An experimental set-up of Eliezer's problem considering a collision between an
electron and a positron, will pose difficulties of the same kind as those
observed in the case of a storage ring. So, even though Eliezer's problem has
a clear meaning in a classical context, where position and momentum have
precise values at any time, quantum effects are unavoidable in the experiment
and, therefore, Lorentz-Dirac equations do not apply.

Why then should we insist on looking for exact solutions of Lorentz-Dirac
equations if they are of little help for an accurate practical description of
the motion of the charge? In our opinion, the study of these equations in
their own context, regardless of experiments or practical applications, is
interesting because their coherence is deeply related to the mass
renormalization problem, which still remains somewhat obscure in classical as
well as quantum theory. It was precisely in Lorentz-Dirac equations where, for
the first time, Dirac introduced the concept of mass renormalization, in order
to deal with the divergent self-energy of a point charge. Thus, it would not
be surprising to find that the violation of the energy conservation law in
these equations is related to the mass renormalization problem.  Now, if the
Lorentz-Dirac equations are incorrect, as we clearly show below, the natural
question that arises is: What are the correct equations of motion for point
charges? This question assumes that the divergence due to the point nature of
the charges can be consistently renormalized with the energy conservation law;
but, is it possible that this program could not be carried out? In order to
answer these and other related questions, Lorentz-Dirac equations must be
studied in their own classical context. The analysis of exact solutions
becomes a powerful tool for this purpose.

Let us now describe our solution. We assume that the motion occurs on the
$X-Y$ plane, clockwise in a circle of radius $a$ centered at the origin. The
two identical particles of charges $e < 0$ and mass $m$, are rotating with
constant angular velocity $\omega$, at the end of a diameter. At time $t$, let
charge 1 be the one located at $x = -a\cos\omega t$, $y = a\sin \omega t$.
Then, the retarded position of charge 2, which determines the field over
charge 1 at time $t$, is defined by the retarded time $t'$, time that also
defines the retarded position of charge 1 respect to charge 2 at time $t$.
This property is due to the symmetry of the motion. Now, let $2\varphi =
\omega t - \omega t'$  be the angle between the
diameters of the actual and the retarded positions.  Then, we have $\varphi =
\beta \cos \varphi$, where $\beta = a \omega /c$. In
particular, we obtain $d\beta/d\varphi > 0$.  Thus, $\beta$ is a strictly
monotonous increasing function of $\varphi$. Therefore, since there exists a
one to one correspondence between $\beta$ and $\varphi$, we will use either of
them to represent the velocity of the charges. Moreover, as $\beta < 1$, then,
$\varphi < \varphi_{{\rm cri}} = 0.739$, and since $\beta > 0$, the angle
$\varphi$ is in the interval $0 < \varphi < \varphi_{{\rm cri}}$.

In order to account for the retarded interaction between the charges (the
second term on the RHS of (1)), we introduce the radial and tangential unitary
vectors to the orbit at the location of charge $1$, which are $\hat{r} =
-\hat{i} \cos \omega t + \hat{j} \sin
\omega t$ and $\hat{t} = \hat{i} \sin \omega t + \hat{j} \cos \omega t$,
respectively. Then, the radial component $E_r = \hat{r} \cdot E$ and the
tangential component $E_t = \hat{t} \cdot E$ of the retarded electric field of
charge 2 at the location of charge 1 at time $t$, are obtained from
Li\'enard-Wiechert's well-known formulas for the field of a point charge. We
get:
\begin{eqnarray}
(4a^2/e) E_r & = & (\cos \varphi + \varphi\sin\varphi)^{-3} \left\{
\left( \varphi \sin 2 \varphi + \cos^2 \varphi \right)
\left( 1 + \varphi^2 \cos 2\varphi \sec^2 \varphi \right)\right.\nonumber \\
& & \left.-2\varphi^2 \left( 1 + \varphi \tan \varphi \right) \cos 2\varphi
\right\},
\end{eqnarray}
\begin{eqnarray}
\left( 4a^2/e\right) E_t & = & \left( \cos \varphi + \varphi
\sin\varphi\right)^{-3} \left\{
\left( \varphi \cos 2 \varphi - \sin \varphi \cos \varphi \right) \right.
\nonumber\\
& &\left. \left( 1 + \varphi^2 \cos 2 \varphi \sec^2 \varphi \right) + 2
\varphi^2 \left( 1 + \varphi \tan \varphi \right) \sin 2 \varphi
\right\}.
\end{eqnarray}

For this type of motion, the magnetic field at the position of each charge has
only one component along axis $Z$, given by:
\begin{equation}
B_z = - \sin \varphi E_r - \cos \varphi E_t.
\end{equation}
The motion under consideration cannot be obtained without external fields. We
will assume here the existence of a time-independent external electric field
tangent to the orbit, and with a fixed magnitude around the orbit circle. In
order to show that this electric field arise from Maxwell's equations, we
consider the ideal sources: a charge density $\rho ({\bf x}, t)$ that vanishes
identically everywhere, and a charge current vector proportional to time $t$
given by ${\bf J} ({\bf x}, t) = -A t \delta (r - b) \hat{\bfvec{\phi}}$,
where we use cylindrical coordinates $(r, \phi, z)$; $\delta$ is the usual
Dirac delta function, and $\hat{\bfvec {\phi}} $ is the unit vector tangent to
the circle $r = $ fixed, pointing in the direction of increasing $\phi$.
Parameters $A$ and $b$ are positive, with radius $b$ of the infinitely long
solenoid smaller than orbit radius $a$.  Then, it is easy to show that for
these sources the fields:
\begin{eqnarray}
{\bf E} ({\bf x}, t) & = & \left( \mu_0/2\right) A \left\{ r - \theta
\left( r - b \right) \left( r - b^2/r\right) \right\} {\hat{\bfvec \phi}}
\nonumber\\
{\bf B} ({\bf x}, t) & = & - \mu_0 At \left\{ 1 - \theta (r - b) \right\}
\hat{k}
\end{eqnarray}
where $\theta (r)$ represents the step function, are solutions of Maxwell's
equations everywhere. The proof comes easier in cylindrical coordinates. This
proof uses either standard rigor with step and delta functions, or the
distribution theory. We need the fields (5) only in region $r > b$, in which
case they reduce to ${\bf E} = \left( \mu_0 Ab^2/2r \right) \hat{{\bfvec
\phi}}$;
\, ${\bf B} = 0$. The tangential electric field
can take any value at a given orbit circle. In addition to the external
electric field, we will also consider an homogeneous time-independent external
magnetic field pointing in the negative direction of axis $Z$. The fourth
component of Lorentz-Dirac equation (1) is then identically satisfied by $z =
0$ at any time. Equation (1) for the component along the $X$ axis, $x =
-a\cos\omega t$, of charge 1, can be combined with the component along the $Y$
axis, $y = a\sin\omega t$, of the same charge, in order to write them in terms
of radial and tangential component as follows:
\begin{equation}
\hat{\bf t} \left\{ e E_t + e E_t^{{\rm ext}} - \left( 2e^2/3a^2 \right)
\beta^3 \gamma^4 \right\} = \hat{\bf r} \left\{ - m a \omega^2
\gamma - e E_r + e \beta B_z + e \beta B_z^{{\rm ext}} \right\},
\end{equation}
where $ E_t^{{\rm ext}}$ is the tangential component of the external
electrical field, $ B_z^{{\rm ext}}$ is the external magnetic field along the
negative axis $Z$, and $\gamma$ is the usual relativistic parameter given by
$(1 - \beta^2)^{ - 1/2}$. It is easy to show that charge 2 also satisfies
equation (6). From this equation, it follows that must have the following
value:
\begin{equation}
e E_t^{{\rm ext}} =\left( e^2 /a^2 \right) f (\varphi),
\end{equation}
where function $f(\varphi)$ is given by:
\begin{equation}
f(\varphi) = (2/3) \varphi^3 \cos \varphi (\cos^2 \varphi -
\varphi^2)^{-2} - (a^2/e)E_t.
\end{equation}
A detailed analysis shows that $f(\varphi) $ is a positive strictly monotonous
increasing function of $\varphi$. For this reason, for a given radius $a$ and
$a$ given velocity $v = a\omega$, equation (7) determines a perfectly
well-defined external field $E_t^{{\rm ext}}$, which in turn imposes a
restriction on parameters $b$ and $A$ in (5).

The value of the external magnetic field $B_z^{{\rm ext}}$ obtained from (6),
is the following:

\begin{eqnarray}
B_z^{{\rm ext}} & = & \left( e/4 ar_0 \right) \left\{ 4 \varphi \left(
\cos^2 \varphi - \varphi^2 \right)^{-1/2} + \left( r_0/a\right)
\varphi^{-1} \left[ \left( 4a^2 /e\right) E_t \varphi \cos \varphi
\right.\right.\nonumber \\
& & +\left. \left.\left( 4a^2 /e \right) E_r \left( \cos \varphi + \varphi
\sin \varphi \right) \right]\right\},
\end{eqnarray}
where $r_0 = e^2/mc^2$.  For given radius and velocity, the above formula
determines $B_z^{{\rm ext}}$ in a unique way.  For $\varphi$ tending to zero
in (9), $B_z^{{\rm ext}}$ tends to infinity, so that the magnetic fields
compensate the repulsive electric field force between the charges.

The first component of (1), representing the energy conservation law, reduces
to:
\begin{equation}
2 e{\bf v} \cdot {\bf E}^{{\rm ext}} = \left( 4e^2c/3a^2\right) \beta^4
\gamma^4 - \left( 4e^2 c/3a^2\right) \beta^4 g(\varphi),
\end{equation}
where $g(\varphi)$ is given by:
\begin{equation}
g(\varphi) = \left (3 a^2/2e \right) \varphi^{-3} \cos^3 \varphi E_t.
\end{equation}
Except for one factor, equation (10) coincides with equation (7). The left
side of equation (10) corresponds to the rate at which the external field
supplies energy to the system composed of the two charges. Now, the kinetic
energy of the charges remains unchanged in this motion. Besides, the energy
stored in the field of the charges remains also unchanged, since the position
of the charges at two different times cannot be distinguished. We emphasize
that we need not pay attention to the precise definition of the concept of
energy stored in the field of the charges, since by the symmetry of the
motion, this energy -whichever it may be- is time-independent. Therefore,
according to the energy conservation law, all the power supplied to the system
of two charges by the external electric field must be radiated away. This
means that the RHS of (10) corresponds to the total rate of radiation emitted
by the two charges. But for known motion of the charges, Maxwell's equations
determine the field generated by the charges, which in turn determines the
total rate of radiation for this motion.  We carried out this somewhat
elaborated calculation in a separate paper \cite{13} . The total rate of
radiation involves a Larmor term for each charge, plus an interference term
due to the field of both charges. The Larmor terms for the two charges are
precisely the first term on the RHS of (10), so its second term must coincide
with the interference radiation term. If we put the interference radiation
term in \cite{13} in the form $- \left( 4e^2 c/3a^2\right) \beta^4
I(\varphi)$; then, function $I(\varphi)$ must coincide with function
$g(\varphi)$ in (11). We plotted these functions in Fig. 1, showing that for
velocities of the charges not close to the speed of light, function
$g(\varphi)$, which comes from Lorentz-Dirac equations, is qualitatively
similar to $I(\varphi)$.  However, these functions are quantitatively
different for any velocity of the charges, and therefore Lorentz-Dirac
equations violate the energy conservation law. We remark that if we consider
only one charge in circular motion, as in the Sokolov-Kolesnikova solution
\cite{10} , the contradiction with the energy conservation
law disappears, since in this case we have to take $E_t = 0$ in (11).

\mbox{}

\hrule

\begin{center}
Figure 1
\end{center}

\hrule

The above discussion suggests that the source of troubles with the energy
conservation law in Lorentz-Dirac equations for more than one charge, comes
 from an inappropriate consideration of the interference radiation between the
charges. So, it seems natural to solve this problem by incorporating the
interference radiation energy in the equations of motion. There are motions
for which this program may succeed, as seems to be the case for the motion
considered herein. Nevertheless, this procedure may not work in all cases.
Charges in close collision may present troubles, because the trick of
oversimplifying the relevance of the fields in the neighborhood of the
singularities of the charges by means of the mass renormalization procedure
may critically disturb the exchange of energy momentum between the
singularities. This disturbance may lead to a violation of the energy
conservation law.

\subsection*{ACKNOWLEDGMENTS}
We wish to thank N. Bralic, C. Dib, K. Lonsted, I. Schmidt, R.  Tabensky, and
D. Walgraef, for useful discussion. We also want to thank FONDECYT whose
support through project 92-0808 has been determinant for the development of
this research.

\newpage

\subsection*{FIGURE CAPTIONS}

Figure 1: The solid line represents function $I (\varphi)$ of the interference
rate of radiation given by $-\left(4e^2c/3a^2\right)
\beta^4 I(\varphi)$.
The dotted line represents the corresponding function $g(\varphi)$ given by
(11), for the interference rate according to Lorentz-Dirac equations.

\mbox{}

}

\end{document}